\documentstyle[12pt]{article}

\begin{document}
\baselineskip 24 pt

\title{Quantum Cryptography Based on Orthogonal States}

\author{Lior Goldenberg and Lev Vaidman}

\date{}

\maketitle

\begin{center}
{\small \em School of Physics and Astronomy \\
Raymond and Beverly Sackler Faculty of Exact Sciences \\
Tel Aviv University, Tel-Aviv 69978, Israel. \\}
\end{center}

\vspace{2cm}

\begin{abstract}
All existing quantum cryptosystems use non-orthogonal states as the carriers
of information. Non-orthogonal states cannot be cloned (duplicated) by an
eavesdropper. In result, any eavesdropping attempt must introduce errors
in the transmission, and therefore, can be detected by the legal users
of the communication channel. Orthogonal states are not used in quantum
cryptography, since they can be faithfully cloned without altering the
transmitted data. In this Letter we present a cryptographic scheme based on
orthogonal states, which also assures the detection of any eavesdropper.
\end{abstract}

\newpage

A basic task in cryptography is exchanging a secret message between two users,
traditionally called Alice and Bob, in a way that no other party can read it.
The only known method to do this in a proven secure way is to use a `one-time
pad', which uses a previously shared secret information called a key. The key,
a sequence of random bits, is used for encrypting the message. The encrypted
message is completely confidential, even if transmitted via a public
communication channel. Thus, the security of any key-based cryptographic
method depends ultimately on the secrecy of the key. All existing classical
key-distribution cryptosystems are not proven to be secure; their secrecy is
based on computational complexity assumptions which sometimes turn out to be
false. In particular, some existing cryptosystems can be broken (in principle)
due to new developments \cite{Shor} in quantum computation. On the other hand,
the secrecy of quantum cryptosystems is guaranteed by the fundamental laws of
quantum mechanics. Any intervention of an eavesdropper, Eve, must leave some
trace which can be detected by the legal users of the communication channel.

In the last years many quantum cryptosystems were suggested. All these schemes
use non-orthogonal states to encode the information. The first
key-distribution scheme was presented by Bennett-Brassard \cite{BB84} in 1984
(a variation of it was already tested experimentally \cite{BBBSS}). In this
scheme Alice transmits single photons polarized along one of four possible
directions,
\mbox{$\updownarrow$}, \mbox{$\leftrightarrow$},
\mbox{$\nearrow\kern-1.02em\swarrow$} or
\mbox{$\searrow\kern-1.02em\nwarrow$}.
The first two are orthogonal in one basis and the other two are orthogonal in
another  basis. The encoding is as follows: Alice chooses, at random, one of
the four states and sends it to Bob. It is agreed that the  states
\mbox{$\leftrightarrow$} and \mbox{$\searrow\kern-1.02em\nwarrow$} stand for
bit value 0, and the  states \mbox{$\updownarrow$} and
\mbox{$\nearrow\kern-1.02em\swarrow$} stand for bit value 1. Bob chooses, also
at random, a basis, {\mbox{$\oplus$}} or \mbox{$\otimes$}, and measures the
polarization in that basis. If Alice and Bob choose the same basis, their
results should be identical. If they choose different bases, their results are
not correlated. By discussing over an insecure classical channel (which cannot
be modified by an eavesdropper), Alice and Bob agree to discard all the cases
where different bases were used (about half of the bits). The result should be
two perfectly correlated strings, unless the transmission was disturbed. Any
eavesdropping attempt must introduce errors in the transmission, since Eve
does not know the polarization of each photon. Whenever Alice and Bob measure
in one basis and Eve in the other basis, the correlation of the strings is
destroyed.

The encoding in quantum cryptography was based on non-orthogonal states since
they cannot be cloned (duplicated) by an eavesdropper. Even an imperfect
cloning attempt (intended to gain partial information) induces errors in the
transmission, therefore it is detectable. In general, any two non-orthogonal
states can be used for quantum cryptography, as shown by Bennett \cite{Ben92}.
On the other hand, orthogonal states can be faithfully cloned, so that Eve can
copy the data without being noticed. For these reasons it is generally
believed that the use of non-orthogonal states is crucial in quantum
cryptography. In this Letter we present a new quantum cryptosystem, in which
data exchange between Alice and Bob is done using two orthogonal states, and
yet, any eavesdropping attempt is detectable.

The security of our scheme is based on two novel ingredients. First, the
orthogonal states sent by Alice are superpositions of two localized
wavepackets. The wavepackets are not sent simultaneously towards Bob, but one
of them is delayed for a fixed time and sent after the other. Second, the
transmission time of each particle is random (and therefore, unknown to Eve).
The tests performed by the users at the end of the communication allows the
detection of an eavesdropper.

Let $|a\rangle$ and $|b\rangle$ be two localized wavepackets, which are sent
from Alice to Bob along two separated channels. We shall take two orthogonal
states $|\Psi_0\rangle$ and $|\Psi_1\rangle$, linear combinations of
$|a\rangle$ and $|b\rangle$, to represent bit value `0' and bit value `1',
respectively:
\begin{eqnarray}
|\Psi_0\rangle  & = & 1 / \sqrt 2 \; (|a\rangle + |b\rangle) ,
\label{Psi0-def} \\
|\Psi_1\rangle  & = & 1 / \sqrt 2 \; (|a\rangle - |b\rangle) .
\label{Psi1-def}
\end{eqnarray}
Alice sends to Bob either $|\Psi_0\rangle$ or $|\Psi_1\rangle$. The two
localized wavepackets, $|a\rangle$ and $|b\rangle$, are not sent together, but
wavepacket $|b\rangle$ is delayed for some time $\tau$. For simplicity, we
choose $\tau$ to be larger than the traveling time of the particles from Alice
to Bob, $\theta$. Thus, $|b\rangle$ starts traveling towards Bob only when
$|a\rangle$ already has reached Bob, such that the two wavepackets are never
found together in the transmission channels.

In order to explain the idea behind the protocol, we shall consider a
particular implementation of our scheme (the discussion assumes a noise-free
transmission). The setup (Fig. \ref{fig1}) consists of a Mach-Zehnder
interferometer with two storage rings, $SR_1$ and $SR_2$, of equal time
delays. Alice can transmit a bit by sending a single particle either from the
source $S_0$ (sending `0') or from the source $S_1$ (sending `1'). The sending
time $t_s$ is random, and it is registered by Alice for later use. The
particle passes through the first beam-splitter $BS_1$ and evolves into a
superposition of two localized wavepackets: $|a\rangle$, moving in the upper
channel and $|b\rangle$, moving in the bottom channel. The particle coming
from $S_0$ evolves into $|\Psi_0\rangle$ and the particle coming from $S_1$
evolves into $|\Psi_1\rangle$. The wavepacket $|b\rangle$ is delayed in the
storage ring $SR_1$ while $|a\rangle$ is moving in the upper channel. When
$|a\rangle$ arrives to the storage ring $SR_2$ at Bob's site, wavepacket
$|b\rangle$ starts moving on the bottom channel towards Bob. During the
flight-time of $|b\rangle$, wavepacket $|a\rangle$ is delayed in $SR_2$.
Finally, the two wavepackets arrive simultaneously to the second beam-splitter
$BS_2$ and interfere. A particle started in the state $|\Psi_0\rangle$ emerges
at the detector $D_0$, and a particle started in the state $|\Psi_1\rangle$
emerges at the detector $D_1$. Bob, detecting the arriving particle, receives
the bit sent by Alice: $D_0$ activated means `0' and $D_1$ activated means
`1'. In addition he registers the receiving time of the particle $t_r$.

Alice and Bob perform two tests (using a classical channel) in order to detect
possible eavesdropping. First, they compare the sending time $t_s$ with the
receiving time $t_r$ for each particle. Since the traveling time is $\theta$
and the delay time is $\tau$, there must be $t_r = t_s + \tau + \theta$.
Second, they look for changes in the data by comparing a portion of the
transmitted bits with the same portion of the received bits. If, for any
checked bit, the timing is not respected or anti-correlated bits are found,
the users learn about the intervention of Eve.

We will show that Eve, which has access to the channels but not to the
sites of Alice and Bob, cannot extract any information without introducing
detectable distortions in the transmission. The data is encoded in the
relative phase between the two wavepackets $|a\rangle$ and
$|b\rangle$. Therefore, the phase must be the same at $t_s$ and at
$t_r$. In addition, the two wavepackets must arrive together to $BS_2$ at
the correct time, otherwise a timing problem occurs. Any operation
performed by Eve must obey these two requirements, or she will be exposed
by the legal users.

Let us consider two times, $t_1$ and $t_2$. At $t_1$ the particle just left
$BS_1$, so it is solely at Alice's site. At $t_2$ the particle is just before
passing through $BS_2$ at Bob's site. If the particle is emitted from $S_0$,
then at $t_1$ its state is $|\Psi_0(t_1)\rangle = 1/ \sqrt 2 \;
(|a(t_1)\rangle + |b(t_1)\rangle)$. If the particle is emitted from $S_1$,
then at $t_1$ its state is $|\Psi_1(t_1)\rangle = 1/ \sqrt 2 \;
(|a(t_1)\rangle - |b(t_1)\rangle)$. In case that nothing disturbs the
transmission (i.e. Eve is not present), the free time-evolution is
\begin{eqnarray}
|\Psi_0(t_1)\rangle & \longrightarrow &
|\Psi_0(t_2)\rangle = 1 / \sqrt 2 \; (|a(t_2)\rangle + |b(t_2)\rangle) ,
\label{frevol-0} \\
|\Psi_1(t_1)\rangle & \longrightarrow &
|\Psi_1(t_2)\rangle = 1 / \sqrt 2 \; (|a(t_2)\rangle - |b(t_2)\rangle) .
\label{frevol-1}
\end{eqnarray}
When Eve is present and she is trying to extract some information without
being detected, the time-evolution must be such that $|\Psi_0(t_1)\rangle$
evolves to $|\Psi_0(t_2)\rangle$ and $|\Psi_1(t_1)\rangle$ evolves to
$|\Psi_1(t_2)\rangle$ (if not, Bob will have a non-zero probability to receive
inverted bits or to receive particles at incorrect times). Thus, the general
form of the evolution from time $t_1$ to time $t_2$ must be:
\begin{eqnarray}
|\Psi_0(t_1)\rangle \, |\Phi(t_1)\rangle & \longrightarrow &
|\Psi_0(t_2)\rangle \, |\Phi_0(t_2)\rangle , \label{Evevol-0} \\
|\Psi_1(t_1)\rangle \, |\Phi(t_1)\rangle & \longrightarrow &
|\Psi_1(t_2)\rangle \, |\Phi_1(t_2)\rangle , \label{Evevol-1}
\end{eqnarray}
where $|\Phi (t)\rangle$ is the  state of some auxiliary system used by Eve
for extracting information. If $|\Phi_0(t_2)\rangle = |\Phi_1(t_2)\rangle$, no
extraction of information is possible.

In protocols which use non-orthogonal quantum states for encryption, the
time-evolution under eavesdropping must have the same form as
eqs.(\ref{Evevol-0}) and (\ref{Evevol-1}). The security of these protocols,
i.e. $|\Phi_0(t_2)\rangle = |\Phi_1(t_2)\rangle$, can be proven using the
unitarity of quantum theory. When Eve is not present, from the free evolution
(eqs.(\ref{frevol-0}) and (\ref{frevol-1})) we get
$\langle\Psi_1(t_1)|\Psi_0(t_1)\rangle =
\langle\Psi_1(t_2)|\Psi_0(t_2)\rangle$. When Eve is present, from
eqs.(\ref{Evevol-0}) and (\ref{Evevol-1}) we get
$\langle\Psi_1(t_1)|\Psi_0(t_1)\rangle = \langle\Psi_1(t_2)|\Psi_0(t_2)\rangle
\, \langle\Phi_1(t_2)|\Phi_0(t_2)\rangle$. Combining these two results we find
$|\Phi_0(t_2)\rangle = |\Phi_1(t_2)\rangle$. With orthogonal states, however,
this proof fails, since $\langle\Psi_1(t_1) |\Psi_0(t_1)\rangle = 0$. For this
reason one might believe that quantum cryptography cannot rely on orthogonal
states.

We shall prove now that our protocol is secure. Using the linearity of quantum
theory, we consider the evolution of a particular superposition of
$|\Psi_0(t_1)\rangle$ and $|\Psi_1(t_1)\rangle$. Consider at time $t_1$ a
particle in the state $|b(t_1)\rangle = 1/ \sqrt 2 \; (|\Psi_0(t_1)\rangle -
|\Psi_1(t_1)\rangle)$. The time-evolution of $|b(t_1)\rangle \,
|\Phi(t_1)\rangle$ is obtained from eqs.(\ref{Evevol-0}) and (\ref{Evevol-1})
(using also eqs.(\ref{frevol-0}) and (\ref{frevol-1})):
\begin{eqnarray}
|b(t_1)\rangle |\Phi(t_1)\rangle \longrightarrow
1/2 \, [|a(t_2)\rangle \, (|\Phi_0(t_2)\rangle \! - \! |\Phi_1(t_2)\rangle) +
|b(t_2)\rangle \, (|\Phi_0(t_2)\rangle \! + \! |\Phi_1(t_2)\rangle)] .
\label{Evevol-b}
\end{eqnarray}
The last equation shows that, unless $|\Phi_0(t_2)\rangle =
|\Phi_1(t_2)\rangle$, there is a non-zero probability to find the particle in
the final state $|a(t_2)\rangle$. This, however, is impossible. A particle in
the state $|a(t_2)\rangle$ is a particle which just emerged from the storage
ring $SR_2$ (there is no other possibility). Since the delay time is $\tau$,
at an earlier time than $t \equiv t_2 - \tau$ the particle had to enter in
Bob's site. At that time, a particle which started in the state
$|b(t_1)\rangle$, as in eq.(\ref{Evevol-b}), is still captured in $SR_1$ at
Alice's site. Such a particle enters in the bottom channel after time $t$, and
then it is too late for Eve to send a dummy particle on the upper channel. She
cannot send that particle at the correct time since she does not know it until
the original wavepacket arrives. Thus, the state $|a(t_2)\rangle$ should not
appear in the right-hand side of eq.(\ref{Evevol-b}), and therefore,
$|\Phi_0(t_2)\rangle = |\Phi_1(t_2)\rangle$. This ends the proof.

We want to emphasize that the sending time cannot be publicly known, otherwise
Eve could apply the following strategy: Using a replica of Alice's setup, she
sends to Bob (at the correct time) a wavepacket $|b\rangle$ of a dummy
particle, while waiting for Alice's particle. Using a replica of Bob's setup,
she measures the later. Depending on the result of the measurement, she places
a phase-shifter in front of the delayed wavepacket $|a\rangle$ of the dummy
particle, in order to adjust the final interference. In this way Eve can
extract the complete information without being exposed.

Since $\tau > \theta$, Eve has no access to $|a\rangle$ and to $|b\rangle$
together at any time. This seems to be a necessary requirement for a secure
protocol, but it is not. If the communication is based on particles moving
at the speed of light, it is enough to demand $\tau > \Delta t$, where
$\Delta t$ is the accuracy of the time measurements of $t_s$ and $t_r$
(assuming very narrow wavepackets). The security in this case is proven in
the same way: the state $|a(t_2)\rangle$ should not appear in
eq.(\ref{Evevol-b}), since Eve gets wavepacket $|b\rangle$ too late for
sending a dummy particle on the upper channel. Moreover, if we arrange a
large distance between the two transmission channels (which requires large
secure users' sites), we can use our procedure even without time delay. Any
attempt of Eve to recombine the wavepackets in order to measure the phase,
introduces an extra flight-time which will be detected by the
users. However, now the security requires that Eve cannot use
faster-than-light particles for eavesdropping. Thus, these versions of the
protocol exceed the limits of non-relativistic quantum mechanics; they
might be classified as ``quantum-relativistic protocols'' with orthogonal
states.

In the previous discussion we have assumed ideal transmission
conditions. In practice, any communication system is restricted by the
limited efficiency of its components. The transmission is distorted by the
noise of the channel, the losses and dark counts of the detectors,
etc. Since errors from different sources are not necessarily
distinguishable, Eve may obtain some information without being detected, as
long as the amount of errors she introduces does not exceed the
noise. Known methods of error correction and privacy amplification
techniques can be included in a practical version of our protocol. The
problems caused by losses and dark counts are automatically solved, due to
the comparison between $t_s$ and $t_r$.

We shall raise some ideas related to the realization of our protocol in the
laboratory. The first essential ingredient, random emission time, can be
achieved very naturally using down-conversion crystal source of pairs of
photons. In this way, the sending time of the photon is registered with very
high efficiency and precision by a detector of the ``idler'' photon. The
second ingredient, the time delay, can be achieved using an optical fiber
loop. Probably, the most difficult part of the proposal is to have a
Mach-Zehnder interferometer with a stable phase difference between its two
(very long) arms. This problem can be avoided using one arm (an optical fiber)
and two orthogonal polarizations as two quantum channels. In this setup
wavepacket $|b\rangle$ leaves Alice's site when it is spatially delayed
relative to wavepacket $|a\rangle$, and with a different polarization. In
Bob's site, wavepacket $|a\rangle$ is delayed and its polarization direction
is rotated, such that the two wavepackets finally interfere correctly.

Since there are some difficulties in an experiment with two polarization
channels, a better way is sending the states with the same polarization,
i.e.  using a single channel. A modification of the setup in
Fig. \ref{fig1} allows the transmission of the wavepackets with the same
polarization, but for the price of wasting a part of the photons
\cite{Tal}. A mirror and a beam-splitter added to Alice's site (after
$SR_1$) can partially recombine the two channels into a single one. Similar
beam-splitter and mirror added to Bob's site (before $SR_2$) can recover
the two channels. As before, the users consider only photons which respect
the timing requirement, but now a part of the sent photons are lost even if
Eve is not present. Half of the photons are lost at Alice's site since they
do not enter into the channel, and half of those which arrive to Bob's site
are lost since they are detected at incorrect times. Thus, only 25\% of the
photons are usable, but this is good enough for key-distribution. The phase
can be preserved more efficiently on a single channel, therefore this
method might be practical for long-range transmission.  One may be tempted
to improve this proposal by introducing a setup which allows Bob to measure
correctly all the transmitted photons. This can be done for the price of
introducing uncertainty in the correlations between the sending and the
receiving time of each photon, but then the method is not appropriate for
our purpose (since Eve has time to get the signal and to resend it without
being detected).

An advantage of using orthogonal states over non-orthogonal states is also
related to the possibility of transmitting signals at long distances:
orthogonal states can be `enhanced' in intermediate stations, as classical
signals are. Measuring a signal many times on the way decreases
dramatically the amount of expected errors, due to the `quantum Zeno
effect'.  The stations, however, have to be secure as the sites of Alice
and Bob are.

Another advantage of our protocol (with two channels) over some other
protocols (for example \cite{Eke91}) is that the bits are not random, but
chosen by Alice, and that all the sent bits can be used. Therefore, the
protocol is not restricted to key-distribution only -- it can be used for
sending the message directly \cite{BB-message}. Of course, Eve can read the
message, but in an error-free channel she will be detected in time if Alice
and Bob test the transmission frequently enough. The direct message
transmission is possible not only on an error-free channel \cite{Tal}. In a
practical case (when noise is present), Alice and Bob agree in advance on
the tolerable error rate and on the degrees of accuracy and secrecy they
want to achieve. In order to transmit a message of some length $n$, Alice
builds a longer string: some extra bits are used for estimating the error
rate (hence, the maximal information leaked to Eve) and some for
redundancy, which is used -- via block-coding -- to derive the $n$-bits
message. The reliability of the $n$-bits message is assured by Shannon's
channel coding theorem, (see \cite{EHPP}). At the end of the transmission,
Alice tells Bob which bits were used for error estimation, and afterwards,
the function used for block-coding. If Bob, estimating the error rate,
detects Eve, he prevents publishing the block-coding function by informing
Alice. Thus, the message is transmitted with an exponentially small
probability of errors and exponentially small information leakage.

Let us conclude with a discussion of the title of our work. Strictly
speaking, the set of all possible states sent by Alice is not a set of
orthogonal states.  Two states corresponding to identical bits, sent at two
very close times, are not orthogonal. However, if the width of the
wavepackets $|a\rangle$ and $|b\rangle$ is small enough, then the measure
of mutual non-orthogonality is negligible. Moreover, we can replace the
random sending times by random discreet sending times, and then, all the
possible sent states will be mutually orthogonal. The previous proof
assures the security of this procedure too. Note also, that in our basic
method (with two channels) all the states corresponding to different bits
are mutually orthogonal, and this is the relevant feature. Indeed, the
issue of mutual orthogonality of just these states is essential for the
security proof of protocols using non-orthogonal states.

The authors thank Tal Mor, Sandu Popescu, David DiVincenzo and Bruno
Huttner for  valuable comments.

\vspace{3cm}
{\bf Figures Caption}

\begin{figure}[h]
\caption{Cryptographic scheme based on a Mach-Zehnder interferometer.
The device consists of two particle sources $S_0$ and $S_1$, a beam-splitter
$BS_1$, two mirrors, two storage rings $SR_1$ and $SR_2$, a beam-splitter
$BS_2$ and two detectors $D_0$ and $D_1$. The device is tuned in such a way
that, if no eavesdropper is present, a particle emitted by $S_0$ ($S_1$) is
finally detected by $D_0$ ($D_1$). \label{fig1}}
\end{figure}


\begin{thebibliography}{99}
\bibitem{BB-message} The same can be done in a modified version of
Bennett-Brassard scheme \cite{BB84}, if Bob can delay his measurements until
he is told the preparation basis of every photon.
\end{thebibliography}
\end{document}